\documentclass[a4paper,11pt]{article}

\usepackage{amssymb}
\usepackage{amsmath}
\usepackage{graphicx}
\usepackage{comment}

\begin{document}

\title{Shell-structure effects on high-pressure Rankine-Hugoniot shock adiabats}

\author{J.C. Pain\footnote{CEA/DIF, B.P. 12, 91680 Bruy\`eres-Le-Ch\^atel Cedex, France, email: jean-christophe.pain@cea.fr}}

\maketitle

\begin{abstract} 
Rankine-Hugoniot shock adiabats are calculated in the pressure range 1 Mbar-10
Gbar with two atomic-structure models: the atom in a spherical cell and the atom
in a jellium of charges. These quantum self-consistent-field models include
shell effects, which have a strong impact on pressure and shock velocity along
the shock adiabat. Comparisons with experimental data are presented and quantum
effects are interpreted in terms of electronic specific heat. A simple
analytical estimate for the maximum compression is proposed, depending on
initial density, atomic weight and atomic number. 
\end{abstract}


\section{\label{sec1} Introduction}

The study of radiative properties of high-energy-density plasmas has many
applications in inertial confinement fusion and astrophysics. It requires
knowledge of the plasma properties in extreme conditions of density and
temperature. For example, the pressure in the center of Jupiter is believed to
reach 100 Mbar and temperature 25 kK, while in the center of white dwarfs
pressure can exceed Gbars at temperatures of 10$^7$ K. In the laboratory,
laser-based shock-wave experiments attain pressures of hundreds of megabars,
obtained via the absorption of an intense laser pulse. The thermodynamics and
the hydrodynamics of hot dense plasmas can not be predicted without a knowledge
of the equation of state (EOS) that describes how a material reacts to pressure.
For instance, the theory of stellar evolution is affected by uncertainties in
EOS. After being predicted, brown dwarfs have recently been observed. Their
internal structure and cooling time depend on the details of the EOS at
densities approaching solid density and at temperatures of a few eV, conditions
obtainable in laser experiments \cite{rem}. Therefore, the need for suitable EOS
of high-energy-density matter becomes crucial. 

A material subjected to a strong shock wave is compressed, heated and ionized.
As the strength of the shock is varied for a fixed initial state, the ensemble
of the pressure-density final states of the material behind the shock, named
shock adiabat or Rankine-Hugoniot (RH) curve, depends on the EOS of the matter,
which must be determined from theory. At intermediate shock pressures, when the
material becomes partially ionized, the EOS depends on the precise
quantum-mechanical state of the matter, \textit{i.e.} on the electronic shell
structure. Therefore, for the past twenty years there is a great interest in the
physics of bound levels in high-energy-density plasmas and quantum
self-consistent-field (QSCF) models \cite{roz} are replacing the older
Thomas-Fermi (TF) approach \cite{fey}. We present RH curves calculated using our
QSCF-based EOS code, ESODE (Equation of State with Orbital Description of
Electrons). The ionic contribution to the EOS is described by an ideal gas with
OCP (One Component Plasma) corrections, and the cold curve (T=0 K isotherm) has
been obtained in most of the cases from Augmented Plane Wave (APW) calculations
\cite{louc}. 

ESODE has two treatments for the thermal electronic contribution to the EOS:
Average atom in a Spherical Cell (ASC) or Average atom in a Jellium of Charges
(AJC). In both cases, bound electrons are treated quantum-mechanically. In the
ASC model, all electrons are confined within a Wigner-Seitz sphere and free
electrons are described in the TF approximation. In the AJC model, bound and
free-electron wavefunctions can extend outside the sphere, where the plasma is
represented by a uniform electronic density (jellium) neutralized by a
continuous background of positive charges, representing ions. The QSCF models
(ASC and AJC) are described in Sec. \ref{sec2}. Shock adiabats calculated from
those models are presented, analyzed and compared to traditional TF model and to
published experimental data in Sec. \ref{sec3}. General features of RH curves,
like the interpretation of shell effects in terms of specific heat, are
discussed in Sec. \ref{sec3} as well. An analytical estimate for the maximum
compression rate is proposed in Sec. \ref{sec4}. Finally, the dependence of
shock velocity on particle velocity is analyzed in Sec. \ref{sec5}.


\section{\label{sec2} The quantum self-consistent-field models}

In the present work, we consider the regime identified as strongly coupled
(non-ideal) plasmas, characterized by a high density and/or a low temperature. In
such plasmas, ions are strongly correlated, electrons are partially degenerate,
and the coupling parameter $\Gamma$, ratio of Coulomb potential energy and
thermal energy, is greater than unity. The main task is to evaluate the thermal
and compressional excitations of the electrons. The determination of the average
electronic charge density relies usually on the Density Functional Theory (DFT),
of which a well-known example is the Thomas-Fermi (TF) model \cite{fey}. It
contains essential features to characterize the material properties in extreme
conditions and is expected to be most reliable at extreme conditions when the
detailed influence of the electronic structure does not play any role. Despite
the omission of quantum structure, TF model typically predicts electronic
densities in broad agreement with detailed approaches. However, as the pressure
or temperature is raised, successive shells of electrons are delocalized and the
effect of such phenomena is represented only in an averaged manner by the TF
model.


\subsection{\label{subsec0}Atom in a Spherical Cell (ASC)}

Atoms in a plasma can be represented by an average atom confined in a
Wigner-Seitz (WS) sphere, whose radius $r_{ws}$ is related to matter density.
Inside the sphere, the electron density has the following form:

\begin{eqnarray}\label{rel1}
n(r)&=&\sum_bf_l(\epsilon_b,\mu)\Big|\psi_b(\vec{r})\Big|^2+\frac{\sqrt{2}(k_BT)
^{3/2}}{\pi^2}[F_{1/2}(-\bar{V}(r),\bar{\mu}-\bar{V}(r),\chi)\nonumber\\
&&+\chi F_{3/2}(-\bar{V}(r),\bar{\mu}-\bar{V}(r),\chi)],
\end{eqnarray}

where 

\begin{equation}
f_l(x,y)=\frac{2(2l+1)}{1+\exp[(x-y)/k_BT]}\;\; \text{and} \;\;
F_{n/2}(a,x,\sigma)=\int_a^{\infty}\frac{y^{n/2}(1+\sigma
y/2)^{1/2}}{1+\exp[y-x]}dy
\end{equation}

are, respectively, the Fermi-Dirac population and the modified Fermi function of
order $n/2$. The first term in Eq. (\ref{rel1}) corresponds to the
contribution of bound electrons to the charge density, while the second term is
the free electron contribution, written in its semi-classical TF form. The
quantity $\epsilon_b$ is the energy of a bound orbital and $\psi_b$ the
associated wavefunction calculated in the Pauli approximation \cite{pau}, in
which only first-order relativistic corrections to the Schr\"odinger equation
are retained. We introduce the notations $\chi=k_BT/E_0$, $E_0$ being the
rest-mass energy of the electron and $\bar{V}(r)=V(r)/(k_BT)$, where $V(r)$ is
the self-consistent potential:

\begin{equation}\label{scf2}
V(r)=-\frac{Z}{r}+\int_0^{r_{ws}}\frac{n(r')}{|\vec{r}-\vec{r'}|}d^3r'+V_{xc}[n],
\end{equation}

$V_{xc}$ being the exchange-correlation contribution, evaluated in the local
density approximation \cite{iye}. Last, the chemical potential $\mu$ is obtained from
the neutrality of the ion sphere:

\begin{equation}\label{scf3}
\int_0^{r_{ws}}n(r)4\pi r^2dr=Z,
\end{equation}

and $\bar{\mu}=\mu/(k_BT)$. Equations (\ref{rel1}), (\ref{scf2}) and
(\ref{scf3}) must be solved self-consistently provided that bound orbitals are
obtained from the Schr\"odinger equation at each step. The electronic pressure
$P_e$ \cite{pai} consists of three contributions, $P_e=P_{b}+P_{f}+P_{xc}$,
where the bound-electron pressure $P_{b}$ is evaluated using the stress-tensor
formula

\begin{equation}\label{presb}
P_b=\sum_{b}\frac{f(\epsilon_b,\mu)}{8\pi
r_{ws}^2(1+\frac{\epsilon_b}{2E_0})}\Big[\Big(\frac{dy_b}{dr}\Big|_{r_{ws}}\Big)
^2+\Big(2\epsilon_b(1+\frac{\epsilon_b}{2E_0})-\frac{1+l+l^2}{r_{ws}^2}\Big)
y_b^2(r_{ws})\Big],
\end{equation}

$y_b$ representing the radial part of the wavefunction $\psi_b$ multiplied by
$r$. The free-electron pressure $P_{f}$ reads

\begin{eqnarray}\label{presf}
P_f&=&\frac{2\sqrt{2}}{3\pi^2}(k_BT)^{5/2}[F_{3/2}(-\bar{V}(r_{ws}),\bar{\mu}-
\bar{V}(r_{ws}),\chi)\nonumber\\
&&+\frac{\chi}{2}F_{5/2}(-\bar{V}(r_{ws}),\bar{\mu}-\bar{V}(r_{ws}),\chi)]
\end{eqnarray}

and $P_{xc}$ is the exchange-correlation pressure evaluated in the local density
approximation \cite{iye}. The choice of the boundary conditions plays a major
role in the expression of pressure due to the fact that the energy of an orbital
depends on the value of the corresponding wavefunction at the boundary of the WS
sphere. In our model, the wavefunction behaves like a decreasing exponential at
the boundary. The internal energy in the AJC model is

\begin{equation}\label{int}
E_e=\sum_iq_i\epsilon_i-\frac{1}{2}\int_0^{r_{ws}}n(r)\int_0^{r_{ws}}
\frac{n(r')}{|\vec{r}-\vec{r'}|}d^3rd^3r'+E_{xc}-\int_0^{r_{ws}}
n(r)V_{xc}(n(r))d^3r,
\end{equation}

where $E_{xc}$ is the exchange-correlation internal energy and $q_i$ the
population of state $i$ (either bound or free). The first term in Eq.
(\ref{int}) can be expressed by

\begin{equation}
\sum_iq_i\epsilon_i=\sum_bf_l(\epsilon_b,\mu)\epsilon_b+E_{f,k}+E_{f,p}.
\end{equation}

$E_{f,p}$ is the potential energy

\begin{eqnarray}
E_{f,p}&=&\frac{\sqrt{2}(k_BT)^{3/2}}{\pi^2}\int_0^{r_{ws}}[F_{1/2}(-\bar{V}(r),
\bar{\mu}-\bar{V}(r),\chi)\nonumber\\
&&+\chi F_{3/2}(-\bar{V}(r),\bar{\mu}-\bar{V}(r),\chi)] \bar{V}(r)d^3r
\end{eqnarray}

and $E_{f,k}$ the kinetic energy

\begin{eqnarray}\label{rel2}
E_{f,k}&=&\frac{\sqrt{2}(k_BT)^{5/2}}{\pi^2}\int_0^{r_{ws}}
[F_{3/2}(-\bar{V}(r),\bar{\mu}-\bar{V}(r),\chi)\nonumber\\
&&+\chi F_{5/2}(-\bar{V}(r),\bar{\mu}-\bar{V}(r),\chi)]d^3r.
\end{eqnarray}

Here only non-relativistic calculations are performed, which correspond to
$E_0\rightarrow\infty$ in Eqs. (\ref{rel1}), (\ref{presb}) and (\ref{presf}), to
make comparisons with the non-relativistic model described in Sec.
\ref{subsec1}. Equations (\ref{rel1}) to (\ref{rel2}) enable one to include
relativistic effects without solving Dirac equation.

\subsection{\label{subsec1}Atom in a Jellium of Charges (AJC)}

In order to go beyond the TF approximate treatment of continuum electron charge
density, it is necessary to use a full quantum-mechanical description of the
continuum states and to consider that both bound and free orbitals can extend
outside the WS sphere, which requires the definition of the environment beyond
the central average ion. One way to address this is the jellium model (or
electron-gas model), \textit{i.e.} a uniform electron density ($-\bar{n}_+$)
neutralized by a positive background $\bar{n}_+$ simulating the ionic charges.
The model relies on a method proposed by Friedel \cite{fri1,fri2} to treat the
electronic structure of an impurity, represented by a spherical potential of
finite range, in an electron gas. It has been further developed by Dagens
\cite{dag1,dag2} and Perrot \cite{per1,per2,per3}. The problem is reduced to
the response of the electronic density to the immersion of a positive charge $Z$
into the jellium. These considerations lead to the following form of the electron
density:

\begin{equation}
n(r)=\sum_bf_l(\epsilon_b,\mu)\Big|\psi_b(\vec{r})\Big|^2+\sum_l\int_0^{\infty}
f_l(\epsilon,\mu)\Big|\psi_{\epsilon,l}(\vec{r})\Big|^2d\epsilon
\end{equation}

with

\begin{equation}
V(r)=-\frac{Z}{r}+\int_0^{R_{\infty}}\frac{n(r')}{|\vec{r}-\vec{r'}|}d^3r'+
V_{xc}(r)-V_{xc}(-\bar{n}_+),
\end{equation}

where $R_{\infty}>>r_a$, $r_a$ being the radius of the cavity. The potential
$V(r)$ is determined, as in the ASC model, in a self-consistent way. The ionic
density is modeled by

\begin{equation}
n_+(r)=\left\{ \begin{array}{ll}
                   0      & \mbox{for $r<r_a$} \\
                   \bar{n}_+ & \mbox{for $r>r_a$.}
		   \end{array}\right.
\end{equation}

The expression of internal energy is \cite{per3}

\begin{equation}
E[n,n_+]=Z ^*[e_k+e_{xc}](-\bar{n}_+)+\Delta E[n,n_+]-\Lambda_TP_1v_a
\end{equation}

with 

\begin{equation}
v_a=\frac{4}{3}\pi r_a^3, \;\;\; \Lambda_T=\frac{\partial \ln Z^*}{\partial 
\ln T}\Big|_{v_a,T} \;\;\; \text{and} \;\;\;
P_1=\frac{\bar{n}_+}{v_a}\int_{r_a}^{R_{\infty}}4\pi r^2V(r)dr. 
\end{equation}

Quantities $e_k$ and $e_{xc}$ are the kinetic and exchange-correlation energies
per free electron and $\Delta E[n,n_+]$ is the immersion energy, \textit{i.e.}
the energy change resulting from the immersion of an ion in the jellium:

\begin{eqnarray}
\Delta
E[n,n_+]&=&\int_0^{R_{\infty}}(e_k[n(r)]-e_k[-\bar{n}_+])d^3r
-\int_0^{R_{\infty}}\frac{Z}{r}(n(r)+n_+(r))d^3r\nonumber\\
&
&+\frac{1}{2}\int_0^{R_{\infty}}\int_0^{R_{\infty}}\frac{n(r)+n_+(r)}{|\vec{r}-
\vec{r'}|}(n(r')+n_+(r'))d^3rd^3r'\nonumber\\
& &+\int_0^{R_{\infty}}(e_{xc}(n(r))-e_{xc}(-\bar{n}_+))d^3r.
\end{eqnarray}

The pressure is obtained by

\begin{equation}\label{pv}
P_e=[P_k+P_{xc}](-\bar{n}_+)+\bar{n}_+U(r_a)+(1-\Lambda_{v_a})P_1, \;\;
\text{where}\;\;\Lambda_{v_a}=\frac{\partial \ln Z^*}{\partial 
\ln{v}}\Big|_{v_a,T}, 
\end{equation}

and

\begin{equation}
U(r)=-\frac{Z}{r}+\frac{1}{r}\int_0^r4\pi
r'^2(n(r')+n_+(r'))dr'+\int_r^{R_{\infty}}4\pi r'(n(r')+n_+(r'))dr'.
\end{equation}

The term $P_k$ is the pressure of a free-electron gas. The pressure in Eq.
(\ref{pv}) is rigorously the derivative of energy with respect to volume. This
is the main difference with Liberman's model \cite{lib,roz}. The major
difficulty with these models is that the average ionization is not well defined
when the outer electrons are delocalized. The only way to make the formalism
variational is to specify the ionization in the AJC model. In other words, the
question is how to define the residual electron density ($-\bar{n}_+$) far away
from the point where the positive charge is introduced into the jellium. A
convenient choice is \cite{per3} $Z^*(v)=Z^*_{TF}(v)$, $Z^*_{TF}$ being the TF
ionization. Then derivatives of ionization with respect to volume and
temperature can be obtained analytically, using the numerical fit proposed by
More \cite{mor2}. At first sight, ASC and AJC models seem to be very different
concerning the modeling of the environment of the atom, isolated and confined in
the ASC model, and immersed in an infinite effective medium in the AJC model.
However, when the electronic structure is calculated in the TF approximation,
these models are equivalent if the jellium density $(-\bar{n}_+)$ is equal to TF
density evaluated at the WS radius. Such a property comes from a variational
principle (minimization of the free energy).

\subsection{Ionic contribution and cold curve}

The adiabatic approximation is used to separate the thermodynamic functions into
electronic and ionic components. The total pressure can be written:

\begin{equation}\label{eos1}
P(\rho,T)=P_c(\rho)+P_i(\rho,T)+P_t(\rho,T),
\end{equation}

where $P_t(\rho,T)=P_e(\rho,T)-P_e(\rho,0)$, is the thermal electronic
contribution to the EOS. The quantity $P_i$ is the ionic pressure and subscript
``c'' characterizes the cold curve obtained from APW \cite{louc} simulations or
using the Vinet \cite{vin} universal EOS. Equation (\ref{eos1}) also holds for
internal energy. In order to take into account non-ideality corrections to the
thermal motion of ions we use an approximation \cite{nik} based on the
calculation of the EOS of a One-Component Plasma (OCP) by the Monte Carlo method
\cite{han}. The ion contribution can be obtained using a formula based on the
Virial theorem:

\begin{equation}\label{ig2}
P_i(\rho,T)=\rho k_BT+\frac{\rho}{3}\Delta E_i(\rho,T),
\end{equation}

where 

\begin{equation}\label{ig3}
\Delta
E_i(\rho,T)=k_BT\;[\Gamma^{3/2}\sum_{i=1}^4\frac{a_i}{(b_i+\Gamma)^{i/2}}-a_1
\Gamma]
\end{equation}

and $E_i(\rho,T)=3k_BT/2+\Delta E_i(\rho,T)$ with $a_1=-0.895929$,
$a_2=0.11340656$, $a_3=-0.90872827$, $a_4=0.11614773$, $b_1=4.666486$,
$b_2=13.675411$, $b_3=1.8905603$ and $b_4=1.0277554$. Such corrections lead to
small differences in the RH curves as shown in Sec. \ref{sec5}.


\section{\label{sec3}Rankine-Hugoniot shock adiabats}


\subsection{\label{subsec3}Definitions and numerical investigations}

The initial state of the plasma is characterized by a density $\rho_0$, a
temperature $T_0$, a pressure $P_0$, and an internal energy $E_0$. $D$ is the
shock velocity, $u$ the matter velocity, $P$ and $E$ are respectively the
pressure and internal energy behind the shock front. These variables obey the
following Rankine-Hugoniot relation \cite{zel}:

\begin{equation}\label{hug}
\frac{1}{2}(P+P_0)\Big(\frac{1}{\rho_0}-\frac{1}{\rho}\Big)=E-E_0,
\end{equation}

The RH curve can be obtained by solving Eq. (\ref{hug}) at each temperature
step. Calculations have been performed for $T\leq$ 6.4 keV for beryllium (Be,
Z=4), aluminum (Al, Z=13), iron (Fe, Z=26) and copper (Cu, Z=29) calculated from
pure Thomas-Fermi EOS, ASC model, and AJC model. The cold curve has been
calculated using Vinet universal EOS \cite{vin} for Be and APW
\cite{louc} method for Al, Fe  and Cu. 
Results from the QSCF models differ strongly from the TF approximation and 
slightly from eachother. In the case of Al, which we use as an
example (see Fig. \ref{fig1}), the difference between the theories appears for
$P\geq$ 3 Mbar and AJC gives higher pressures than the other models for
2$\leq\eta\leq$ 3.5, where $\eta=\rho/\rho_0$ is the compression rate. For
$\eta\geq$ 3.5, all QSCF models give lower pressures than TF model. 


\subsection{Quantum orbital effects}

\begin{figure}
\begin{center}
\includegraphics[width=8.6cm,trim=0 0 0 -50]{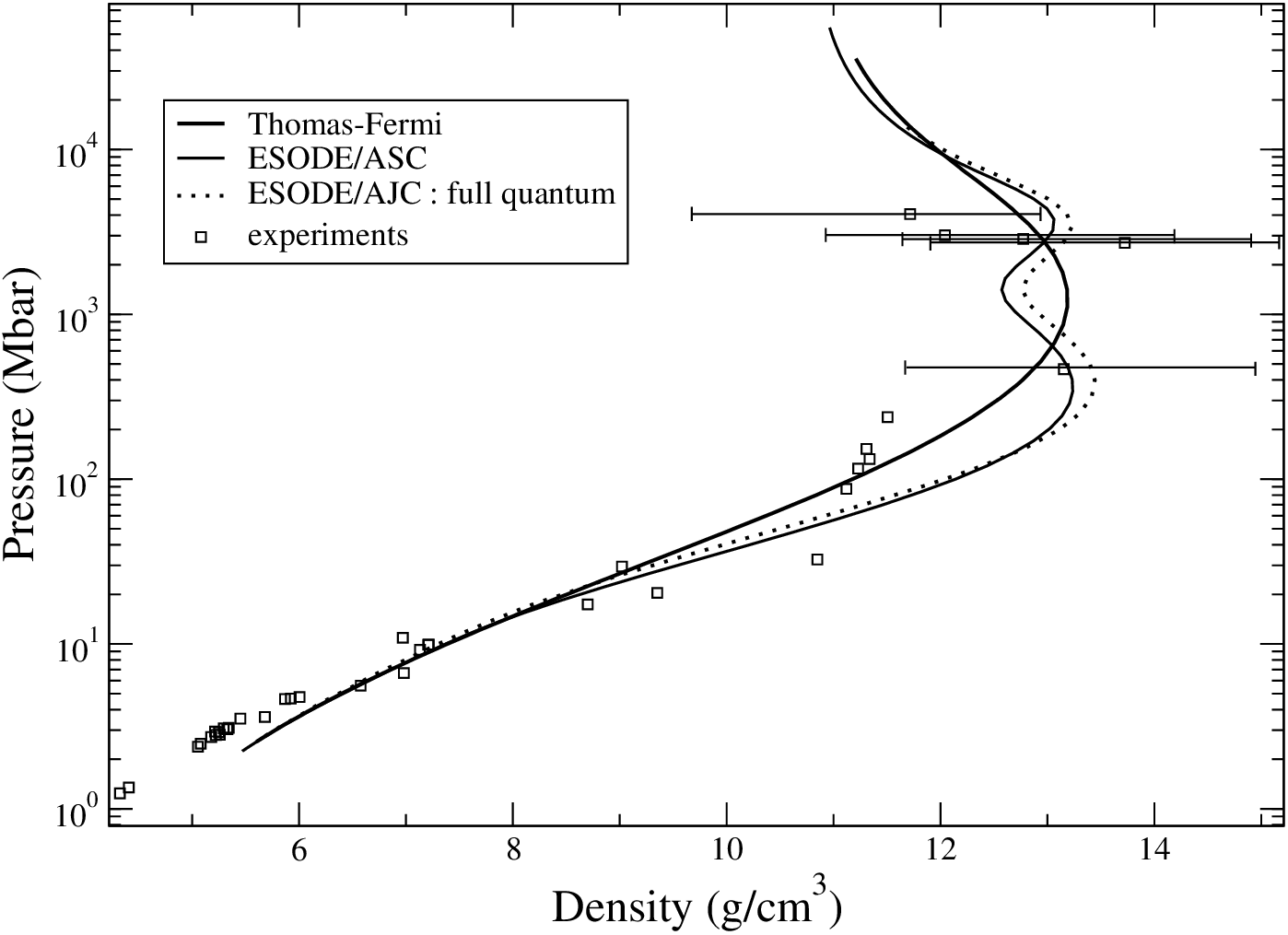}
\end{center}
\caption{Rankine-Hugoniot curves for Al. $\rho_0=$2.70 g/cm$^3$.}
\label{fig1}
\end{figure}

\begin{figure}
\begin{center}
\includegraphics[width=8.6cm,trim=0 0 0 -50]{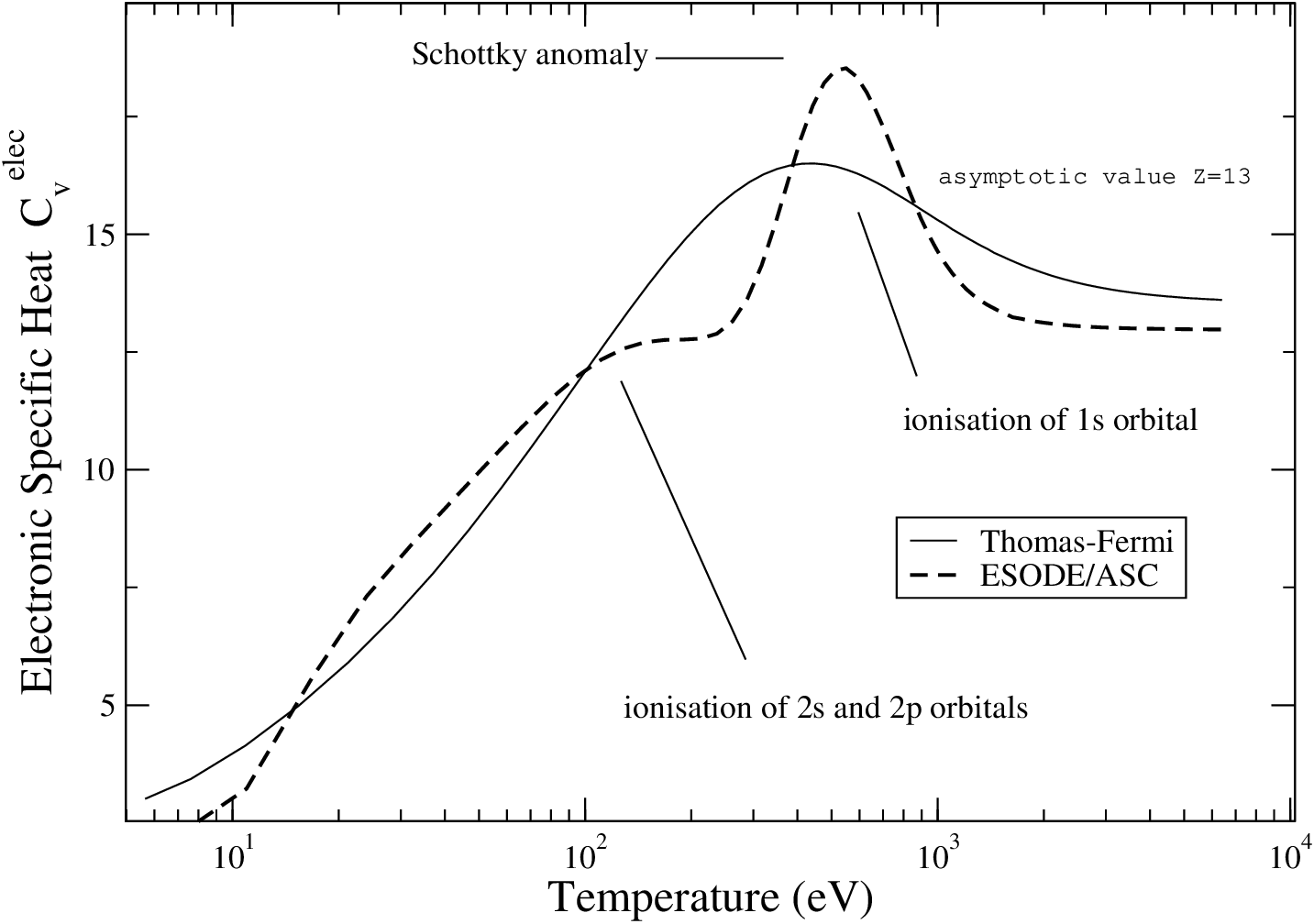}
\end{center}
\caption{Electronic specific heat for Al.}
\label{fig2}
\end{figure}

Our models emphasize the thermodynamic domain where RH curves strongly depend on
the electronic structure, \textit{i.e.} beyond four times solid density where
the shoulders (double in the case of Al, Fig. \ref{fig1}) correspond to
ionization of successive shells. These shoulders can be explained by the
competition between the release of internal energy stored in the shells and the
free-electron pressure. When ionization begins, the energy of the shock
depopulates the relevant shells and the material is very compressive. However,
the pressure due to the free electrons eventually dominates and the material
becomes less compressive. Both models show compression maxima in the range
$5\rho_0-6\rho_0$. In this region, the electrons from the ionic cores are being
ionized. The shock density increases beyond the infinite pressure of $4\rho_0$
in the electron ionization region. As ionization is completed, the plasma
approaches an ideal gas of nuclei and electrons and the density approaches the
fourfold density $4\rho_0$. For Be, all the models yield a single density
maximum, corresponding to the ionization of the K electron shell. For Al (Fig.
\ref{fig1}), there are two density maxima corresponding to the K and L electron
shells, each followed by density decreases. For Fe and Cu there are density
maxima or inflexions corresponding to the K, L and M electron shells. The
L-shell ionization feature gives the largest density increase. In the case of
Cu, the ASC model exhibits a kind of discontinuity around 40 g/cm$^3$, due to
pressure ionization of $2p$ orbital. Such a sharp increase of pressure does not
exist in the AJC model, because this model is consistent from the thermodynamic
point of view. This is due to the fact that the pressure is rigorously obtained
as a derivative of the free energy, and shape resonances are carefully taken
into account in the quantum treatment of free electrons. Such features lead to a
continuous disappearing of a bound state into the continuum, which is a suitable
treatment of pressure ionization. In fact, the non-monotonic character of
thermodynamic variables stems from the eigen-energies of the orbitals
themselves, which exhibit the oscillations as well. The
first density for which decompression occurs is named ``turnaround'' point. The
pressure differences from the quantum mechanical theory can be explained by
examining the electronic heat capacity per particle predicted by the two
theories along the RH path. At low temperature, the electronic heat capacity
depends on the number of electrons that can be excited around the Fermi energy
and the TF theory predicts a smooth increase since the density of states in this
model is a monotonic function of energy. Therefore, the electronic specific heat
per particle
 
\begin{equation}
\tilde{C}_v^{el}=\frac{C_v^{el}}{3k_B/2} \;\;\; \text{where} \;\;\;
C_v^{el}=\frac{\partial [E(\rho,T)-E_i(\rho,T)]}{\partial T}
\end{equation}

displays the signature of the ionization of successive shells. Figure \ref{fig2}
represents $\tilde{C}_v^{el}$ along the RH curve in the TF and in the ASC models
for Al. Both theories show the effect of the coulomb attractive potential of the
nucleus binding the electrons, represented by the peak around 300 eV for TF
theory and 100 eV for ASC model. After the first reduction in density on the RH
curve, there are 11 free electrons and 2 bound electrons remaining in the $1s$
orbital (K shell), which is far away from the energy zero (1.5 keV at 100 eV).
As long as temperature is not sufficient to ionize those two electrons, the
specific heat tends to an asymptote corresponding to an ideal gas of 11
independent particles. When both $1s$ electrons are ionized (after a
``threshold'' temperature), there is a sudden break in the specific heat, which
tends to an ideal gas of 13 independent particles. This phenomenon is a kind of
``Schottky anomaly'', \textit{i.e.} an enhancement in the specific heat (see
Fig. \ref{fig2}). This effect is not as important for the $2s$ and $2p$ bound
states (L shell), since their energy levels are not as far from the continuum (a
few tenth of eV). We note that the same phenomenon occurs for Fe; in that case, after two ``Schottky anomalies'', the
electronic part of the specific heat tends to an ideal gas of 26 electrons.

\subsection{Comparisons with experimental data}\label{tezer}

Experimental data concerning Be, Al, Fe and Cu \cite{rus} have been collected
for comparison with the EOS models presented. Several experimental methods have
been used to generate well-defined shock states: gas guns for pressures up to 5
Mbar, explosive-driven spherical implosions, laser-driven plane waves for
generating shocks up to 10 Mbar, and through underground nuclear explosions
where pressures of a few Gbar can be attained. Superimposing these data should
yield a single smooth RH curve. The maximum pressures reached in the experiments
are: 18 Mbar for Be, 4000 Mbar for Al, 191 Mbar for Fe and 204 Mbar for Cu. It
is clear that there is difficulty discriminating between models, since there are
very few available data for the region of interest, \textit{i.e.}, above 100 Mbar). The
error bars associated with the highest pressures for Al are too large to provide
insight into the existence of shell effects. Analysis of the computational
results shows that the deviation from the experimental points of Al can not be
explained only by shell effects. In the region where most of the experimental
points are available, the role of the cold component is important, while shell
effects begin to play a significant role after the matter is compressed, which
occurs near the limiting compression $\eta\approx 4$, and heating begins. For Fe
and Cu, results from AJC model are found to be in better agreement with
experimental points than the hybrid model ASC. Finally, we note that it is very
difficult to relate gas gun measurements with a discussion of shell effects as
the pressures are too low to exhibit these effect.

\section{\label{sec4} Evaluation of maximum compression}

The maximum compression rate depends on the choice of the boundary conditions of
the wavefunctions described in Sec. \ref{subsec0}. Furthermore, it depends on
the calculation of the ionic part: ideal gas (IG) with or without OCP
corrections (Eqs. (\ref{ig2}-\ref{ig3})). For instance, the maximum rate for Al
in the ASC model with IG is $4.901$ and with IG and OCP corrections $4.931$. OCP corrections
systematically increase the maximum compression rate. As can be checked in Fig.
\ref{fig1} the maximum compression attainable by a single shock is larger than 4
and occurs at finite pressure. This phenomenon is due to the draining of
internal energy in internal degrees of freedom, \textit{via} ionization. An
analytical expression for the maximum compression attainable by a single shock
in any material from any initial state, except those with gaseous densities, can
be formulated. The total energy can be written as the sum of kinetic and
potential energies $E_k$ and $E_p$. Neglecting exchange-correlation
contribution, the virial theorem enables one to relate pressure, kinetic energy
and potential energy $3P/\rho=2E_k+E_p$. The compression rate $\eta=\rho/\rho_0$
for the standard RH curve, $P_0=0$, $\rho_0$ solid density and $T_0$=300 K, can
be written

\begin{equation}
\eta=4+\frac{3}{1+2\frac{\delta E_k}{\delta E_p}} \;\; \text{with} \;\; \delta
E_k=E_k-E_{k_0} \;\; \text{and} \;\; \delta E_p=E_p-E_{p_0}.
\end{equation}

At high compression, assuming that $E_k>>E_{k_0}$ and that all the electrons
have been ionized and have a kinetic energy equal to the Fermi energy, one can
write:

\begin{equation}\label{approx1}
\delta E_k=Z\frac{1}{2}(3\pi^2Z\rho\frac{N_A}{A})^{2/3}a_0^2,
\end{equation}

where $a_0$=52.9177208319 10$^{-10}$ cm is the Bohr radius, $\rho$ is in
g/cm$^3$, and $A$ in g. Next, at high compression the excess potential energy
can be estimated as the Coulomb interaction energy of two ionic spheres

\begin{equation}\label{approx2}
\delta E_p=\frac{1}{2}\frac{Z^2}{r_{ws}} \;\;\text{and}\;\;
r_{ws}=\Big[\frac{3A}{4\pi N_Aa_0^3}\Big]^{1/3}\rho^{-1/3},
\end{equation}

where $r_{ws}$ is the WS radius equal to the cavity radius $r_a$ and we set
$4\pi\epsilon_0=1$. Therefore, using Eqs. (\ref{approx1}) and (\ref{approx2}),
which are relevant for a strongly coupled gas of degenerate electrons, the
maximum compression rate $\eta_m$ obeys 

\begin{equation}
\eta_m=4+\frac{3}{1+\zeta(\rho_0,Z,A)\eta_m^{1/3}},
\end{equation}

with

\begin{equation}
\zeta(\rho_0,Z,A)=3\pi(2 N_A)^{1/3}a_0\Big(\frac{\rho_0}{ZA}\Big)^{1/3}.
\end{equation}

\begin{figure}
\begin{center}
\includegraphics[width=8.6cm,trim=0 0 0 -50]{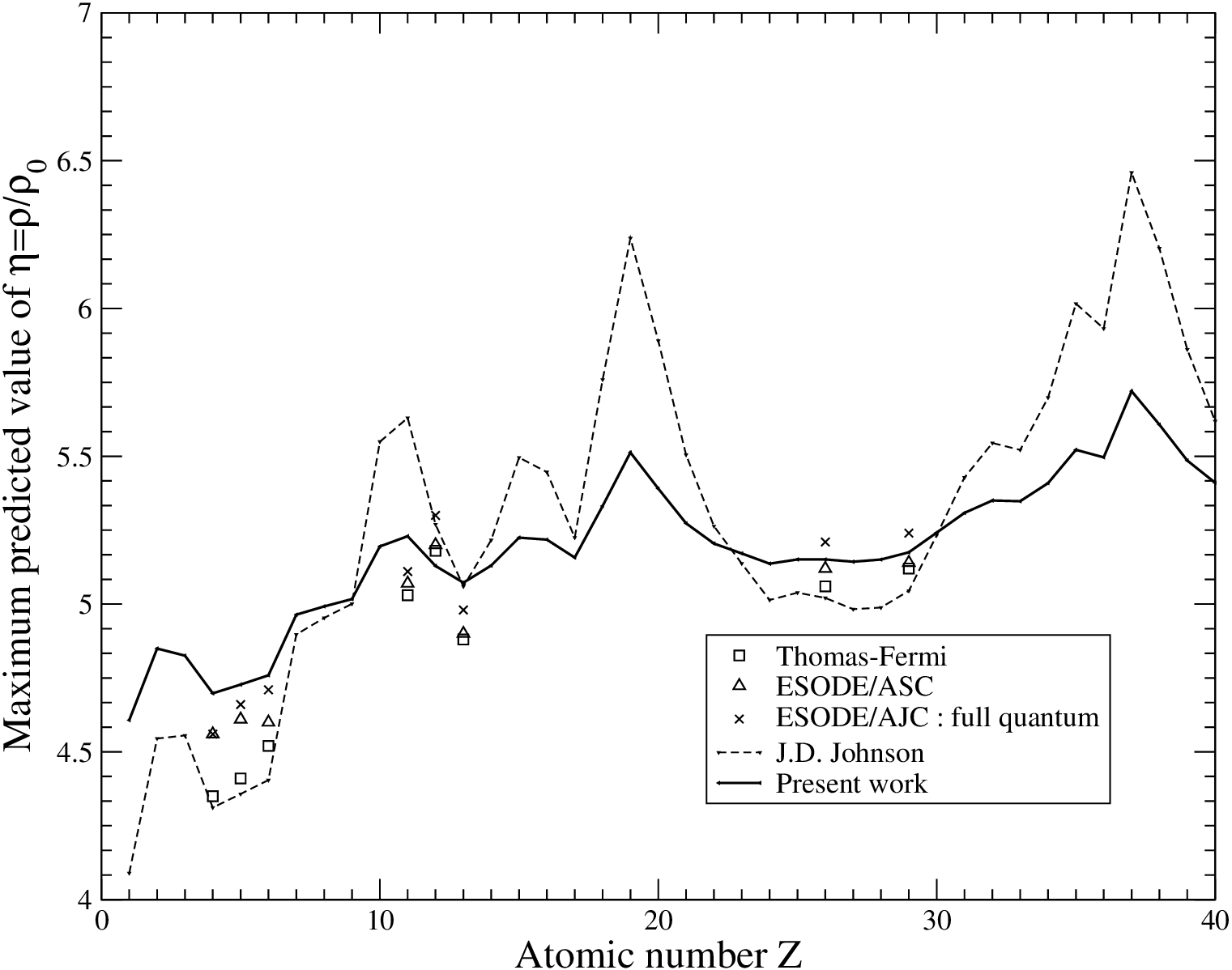}
\end{center}
\caption{Maximum compression rate obtained from \cite{joh}, from our model and
compared to the maximum compression observed from a TF, ASC and AJC calculations
for Be, B, C, Na, Mg, Al, Fe and Cu.}
\label{fig3}
\end{figure}

The solution is:

\begin{equation}\label{eta}
\eta_m=\Big[\frac{-1-(-1)^{\theta(\zeta-\zeta_c)}2\zeta\sqrt{h(\zeta)}}{{4\zeta}
}+\frac{1}{2}\sqrt{\frac{3}{4\zeta^2}-h(\zeta)-(-1)^{\theta(\zeta-\zeta_c)}\frac
{32\zeta^3-1}{4\zeta^3\sqrt{h(\zeta)}}}\Big]^3 
\end{equation}

where $\zeta_c=0.314980262473$, $\theta$ is Heaviside function,

\begin{equation}
h(\zeta)=\frac{1}{4\zeta^2}-\frac{2^{10/3}}{\Delta^{1/3}(\zeta)}+
\frac{\Delta^{1/3}(\zeta)}{2^{1/3}\zeta},
\end{equation}

and $\Delta(\zeta)=-7+16\zeta^3+\sqrt{49+1824\zeta^3+256\zeta^6}$. Figure
\ref{fig3} confirms the fact that the maximum compression is always smaller than
7 \cite{joh}, and strongly dependent on the density $\rho_0$. Note that Eq.
(\ref{eta}) can not be applied for gaseous ambient elements. Neglecting cohesive
and dissociation energies, and using fits for the total ionization energies,
Johnson \cite{joh} has proposed an analytical formula for the maximum
compression rate. It seems difficult, however, to discriminate between his
approach and ours with the values displayed in Fig. \ref{fig3} and calculated
with our QSCF models, since the disparity of values is as large as the
difference between the two models. However, the present calculation does not
rely on a particular form of the EOS, as in Ref. \cite{joh}. On the contrary,
here the maximum compression predicted by our model is higher for Fe than for
Al, which is consistent with the results presented in Ref. \cite{roz}. However,
in Ref. \cite{roz} the maximum compression seems to increase with $Z$; according
to Johnson's model and ours, this is not true \textit{stricto sensu} but is only
a global trend.


\section{\label{sec5} Shock and particle velocities}

The particle velocity and shock velocity are:

\begin{equation}\label{vel1}
u=\sqrt{\frac{(\rho-\rho_0)(P-P_0)}{\rho_0\rho}}\;\; \text{and} \;\;
D=\sqrt{\frac{\rho(P-P_0)}{\rho_0(\rho-\rho_0)}},
\end{equation}

\begin{figure}
\begin{center}
\includegraphics[width=8.6cm]{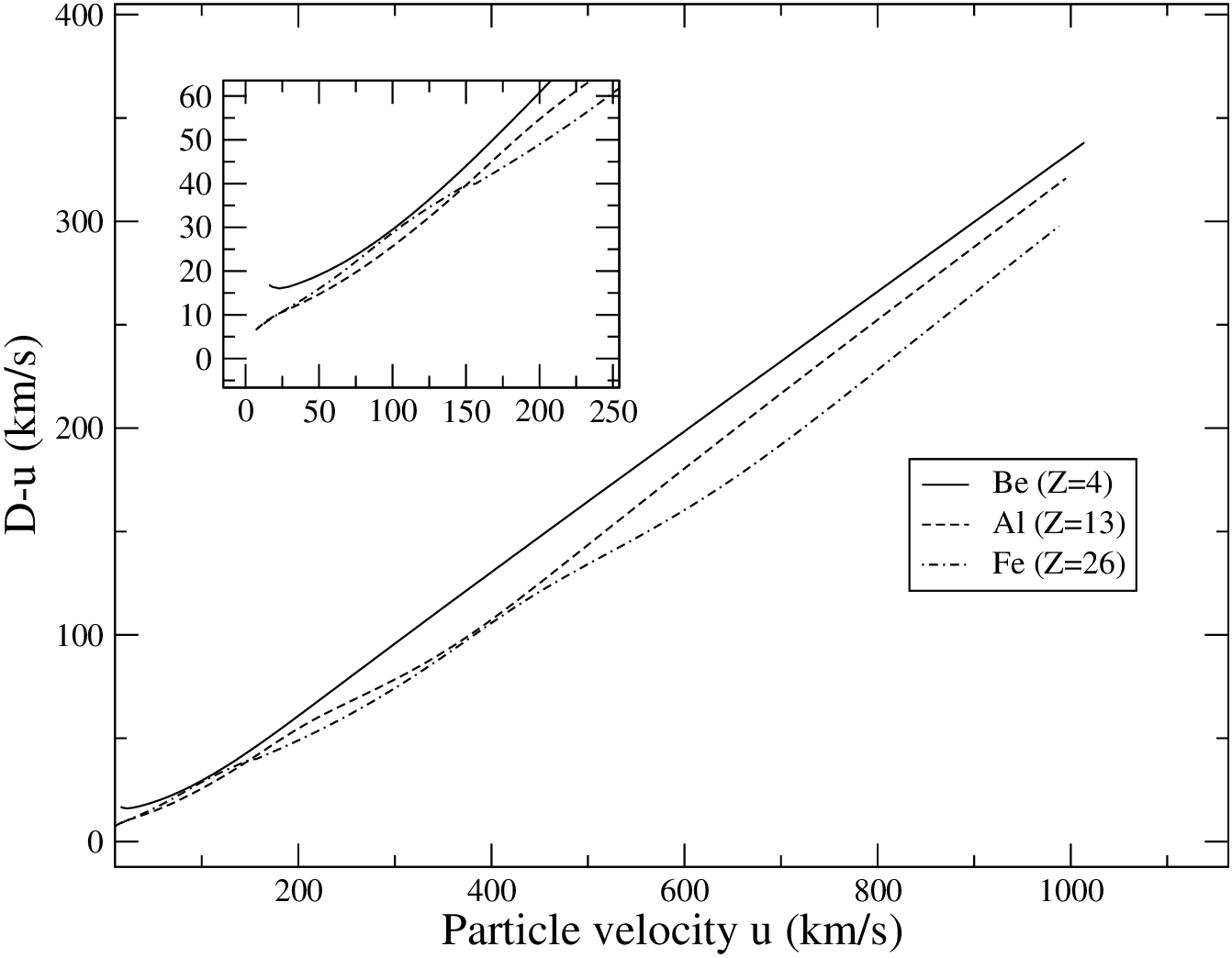}
\end{center}
\caption{$(D-u)$ versus particle velocity $u$ for Be, Al and Fe.}
\label{fig4}
\end{figure}

Expressions in Eq. (\ref{vel1}) are generic as they do not involve, \textit{a
priori}, any explicit relation $D(u)$. For metals, it is often assumed that the
relation is quasi-linear. However, we find that the $(D,u)$ relationship is
almost linear over a wide range of densities, except near $u = 0$ where we could
not perform the calculation. It is easy to check from expressions in Eq.
(\ref{vel1}) that the slope should be $\sim 4/3$, corresponding to the ideal
gas. However, when looking carefully at the first and second derivatives of
shock velocity versus particle velocity, one finds that the behaviour of shock
velocity is more complicated, and that there are some oscillations, reflecting
the shell structure as well, and inflexions points. Deviations from linearity,
indicated by curvature, are more obvious when one investigates the relationship
$(D-u)$ versus $u$, which is illustrated in Fig. \ref{fig4}. The amplitude of
the oscillations in the $(D,u)$ relationship is very small, which is not the
case of $(P,\rho)$ relationship.


\section{\label{sec6} Conclusion}

Shock waves generating finite-temperature dense matter make possible the
generation in laboratory experiments of extremely high energy densities typical
of matter in the few first microseconds after the creation of the universe and
for such astrophysical objects as stars and giant planets. The physical
information obtained from these experiments extends the basic knowledge of
physical properties of these systems to pressures nine orders of magnitude
higher than found, \emph{e.g.}, in our atmosphere. We proposed a qualitative and
quantitative study of quantum orbital effects on the principal shock adiabat for
different elements, for two quantum self-consistent-field models: the atom in a
spherical cell (ASC) and the atom in a jellium of charges (AJC). Quantum orbital
effects lead to oscillations corresponding to the ionization of successive
orbitals that are also visible in the electronic specific heat, and in the
energies of the orbitals. The AJC model provides a better treatment of pressure
ionization, since it relies on a full quantum treatment of the electrons. An
estimate of the maximum compression, giving realistic values, has been
proposed. 

The next step will be to test whether the oscillations still exist ``beyond''
the Average Atom model, \textit{i.e.}, when a variety of electronic
configurations is taken into account \cite{pai}. It remains a difficulty in the
existing models to represent, in a simple and suitable way, the influence of the
plasma environment on a specific ion. Indeed, as this environment fluctuates,
the number, the localization in space and the structure of neighboring ions will
change drastically. Therefore we will in the future represent the ionic
environment including radial distribution functions, thereby going beyond the
adiabatic approximation.

\clearpage

\end{document}